\documentclass[runningheads]{llncs}
\usepackage[utf8]{inputenc}
\usepackage[T1]{fontenc}
\usepackage{graphicx}
\usepackage{booktabs}
\usepackage{array}
\usepackage{tabularx}
\usepackage{longtable}
\usepackage{ragged2e}
\usepackage{amsmath}
\usepackage{amssymb}
\usepackage{textcomp}
\usepackage[hidelinks]{hyperref}
\usepackage{xurl}

\hypersetup{pdfauthor={Rohit Sharma, Pavani Ayinampudi, Aditya B.M.V., Jinal Gupta, Prakash Hegade, Sakshi Sharma, Meenakshi V, SRS Iyengar},pdftitle={Characterizing Questioning Patterns and Student Engagement Through Contextual Analysis of Real-Time Classroom Interactions}}

\begin{document}
\title{Characterizing Questioning Patterns and Student Engagement Through Contextual Analysis of Real-Time Classroom Interactions}
\titlerunning{Questioning Patterns and Student Engagement in Real-Time Polls}
\author{Rohit Sharma\inst{1} \and
Pavani Ayinampudi\inst{2} \and
Aditya B.M.V.\inst{2} \and
Jinal Gupta\inst{2} \and
Prakash Hegade\inst{2} \and
Sakshi Sharma\inst{1} \and
Meenakshi V\inst{1} \and
SRS Iyengar\inst{1}}
\authorrunning{R. Sharma et al.}
\institute{Indian Institute of Technology Ropar, Rupnagar, Punjab 140001, India
\and
ANNAM.AI, Rupnagar, Punjab, India}
\maketitle
\begin{abstract}
Real-time classroom polling is now routine, yet the data it produces is usually read narrowly, as a correctness score or a headcount. Such readings say little about what a poll is doing within a lecture or how it shapes engagement. This is particularly relevant for short-response formats such as True/False, where the same question format can be used to test recall, check comprehension, or direct students' attention to a deliberately misleading statement. This study asks whether a poll's answer and instructional function can be determined by reading it against its lecture transcript, what cognitive levels of Bloom's taxonomy and instructional-function clusters the corpus contains, and how student engagement relates to answering correctly. We analyse a naturalistic corpus of 47 live sessions over 39 days, comprising 604 poll questions and 340,668 responses from 2,807 learners, most items True/False, read against time-aligned lecture transcripts and attendance. Reading each poll in context proves essential: the answer to 89\% of polls is locatable in the lecture, and a recurring attention-checking device is visible only through context. Questioning is overwhelmingly lower-order and falls into seven instructional functions, and a poll's response follows its function rather than its wording. Engagement is broad but concentrated, and the class majority answers correctly 88.5\% of the time, though a small set of high-consensus yet incorrect answers cannot be detected by agreement alone. An independent survey of 579 students agrees on what the polls are and on their participation, but reveals a gap between perception and reality: students cannot judge their own correctness, and the polls they find hardest are not those they answer worst.
\keywords{Bloom's taxonomy \and Classroom polling \and Instructional function \and Learning analytics \and Real-time formative assessment \and Student engagement}
\end{abstract}

\section{Introduction}

Classroom instruction involves continuous interaction between instructors and students. Instructors present ideas, explain concepts, monitor students' understanding, and adapt their teaching as the session progresses \cite{blackwiliam1998}. Students contribute to this process through their participation, responses, and interactions during the session. These interactions allow instructors to observe how students are engaging with the material and how the instruction is progressing. When such interactions are recorded across multiple classes and sessions, they generate a large amount of data that can contain recurring patterns in instructional activity and student engagement. Examining such a large amount of data manually can make it difficult to identify these patterns across sessions. Digital tools can make this analysis feasible by allowing classroom interactions to be organised and analysed across many sessions \cite{ferguson2012}.

One of the ways of classroom interaction is real-time polling. These polls allow instructors to ask questions and collect responses during instruction. Instructors increasingly use clickers, web-based response systems, and polling features built into video-conferencing platforms to collect student responses almost instantaneously \cite{caldwell2007,kaylesage2009,hunsu2016}. The resulting records are commonly summarised through performance, measured by the proportion of learners who answer correctly, and participation, measured by the proportion who submit a response \cite{fiesmarshall2006,fredricks2004}. Both measures provide useful information about what happened after a question was asked, but they say little about the purpose the question served within the lecture or the kind of thinking it required from students.

Questions asked during a lecture can require students to do different things even when they produce the same type of response. A True/False question may ask students to recall a fact, check an explanation, reinforce a recently introduced idea, or pay attention to a deliberate change in a statement. A response to each question can therefore have a different instructional meaning even when the response options are identical. Looking only at whether students responded or answered correctly cannot distinguish between these cases \cite{momsen2010,chin2007}. The question, therefore, carries information about its instructional purpose that is not captured by response counts or correctness scores.

Understanding the instructional purpose of a question also requires considering the context in which it is asked. Its meaning can depend on what the instructor has explained before it, the examples used to develop the topic, and the point at which it appears in the lecture. A True/False question may check recall when it follows a factual explanation, but may check understanding when it follows a conceptual explanation. Questions can also prompt retrieval, redirect attention, check understanding, or provide feedback during instruction \cite{roedigerkarpicke2006,karpickeroediger2008,bunce2010,szpunar2013,blackwiliam1998,hattietimperley2007}. Reading the question together with its surrounding lecture can therefore reveal why it was asked and what students were expected to consider at that point.

A poll can therefore be viewed as part of a larger classroom interaction consisting of the lecture context, the question, and the responses it generates. Existing analyses of classroom polling rarely consider these elements together, leaving the relationship between the instructional purpose of a poll and the responses it receives less clearly understood. In this study, we use this interaction as the basis for examining questioning patterns and student engagement in real-time classroom polls. We analyse 604 real-time polls from 47 live sessions using time-aligned poll records, lecture transcripts, and attendance data.

\section{Background Study}

Research on classroom polling has examined its effects on learning, participation, attention, and feedback for more than two decades. Reviews by Fies and Marshall \cite{fiesmarshall2006}, Caldwell \cite{caldwell2007}, and Kay and LeSage \cite{kaylesage2009} report benefits such as increased participation, greater attention, and timely feedback, while also describing the practical challenges of using polling effectively. A meta-analysis by Hunsu et al. \cite{hunsu2016}, covering more than fifty studies, found that these benefits were real but generally modest and depended on how the polling tool was used. Across this literature, outcomes appear to depend more on the pedagogy surrounding polling than on the technology itself \cite{fiesmarshall2006}. This finding is consistent with broader evidence that active-learning approaches generally produce better learning outcomes than passive lectures in science and engineering \cite{hake1998,freeman2014}. Less attention has been given to what the response record itself contains and to how student responses vary across individual polls.

Attention to the question itself has a long precedent in Peer Instruction, where conceptual questions became a central part of classroom interaction and learning. Studies by Mazur \cite{mazur1997}, Crouch and Mazur \cite{crouchmazur2001}, and Smith et al. \cite{smith2009} found that discussing a question with peers between successive votes can improve students' responses. This work also emphasises the purpose and quality of the question. A useful question is one that requires students to reason and can reveal common misconceptions, rather than one that is judged only by its wording. A similar view appears in formative-assessment research \cite{blackwiliam1998,blackwiliam2009}, as well as in work that distinguishes types of feedback by their instructional function, such as confirming an understanding, identifying a difficulty, or guiding students toward a different approach \cite{hattietimperley2007,shute2008}. Together, these studies suggest that classroom questions and responses can be understood in terms of the instructional function they serve, rather than through correctness alone.

Cognitive psychology provides evidence that questions can support learning during lectures through retrieval and attention. Answering questions can strengthen retention more effectively than restudying the same material, while brief questions during instruction can reduce mind-wandering and improve retention of subsequent material \cite{roedigerkarpicke2006,karpickeroediger2008,bunce2010,risko2012,szpunar2013}.

Two further areas of research help explain what polls require from students and how students respond to them. The original Bloom's taxonomy \cite{bloom1956} and its revised, verb-oriented version \cite{andersonkrathwohl2001,krathwohl2002} provide two ways to characterise the cognitive demand of a question. Previous studies that code assessment items have repeatedly found a skew toward lower-order cognitive processes \cite{zheng2008,momsen2010}. Research on teacher questioning also classifies the questions instructors ask according to their cognitive level and communicative function \cite{redfieldrousseau1981,chin2007,graesserperson1994}. These approaches provide established ways to describe the cognitive demands of classroom questions, but they generally assess questions as individual items rather than in the instructional context in which they are used. A separate area of learning-analytics research has consistently found that online participation is uneven, with a small group of highly active participants alongside a larger group of less active participants \cite{kizilcec2013,nonneckepreece2000,sun2014,henrie2015}. Participation is also commonly treated as a form of behavioural engagement \cite{fredricks2004}, but such analyses often focus on overall participation rather than variation across individual classroom questions. Taken together, these studies provide ways to describe the cognitive demand of questions and patterns of student participation, but they generally examine these dimensions separately rather than as parts of the same classroom interaction. Less is known about how the instructional purpose of an individual poll relates to the responses it receives when the question is interpreted within the lecture in which it occurs. There is also limited evidence connecting such contextual interpretations of poll responses with students' own reports of their experience.

\section{Methodology and Methods}
\label{sec:method}

\subsection{Research Questions}
Our main research question asks what a transcript-grounded reading of real-time polls can reveal about how instructors question and how students engage and answer, beyond what correctness scores or participation counts capture. We address this question through three sub-questions:

\begin{list}{}{%
  \setlength{\leftmargin}{2.6em}%
  \setlength{\labelwidth}{2.1em}%
  \setlength{\labelsep}{0.5em}%
  \setlength{\itemsep}{2pt}%
  \setlength{\parsep}{0pt}%
  \setlength{\topsep}{4pt}%
  \renewcommand{\makelabel}[1]{\textbf{#1}\hfil}%
}
\item[RQ1.] Can a poll's answer and instructional function be determined by reading it in its lecture context?
\item[RQ2.] What cognitive levels and instructional functions does the corpus of polls contain?
\item[RQ3.] How is student engagement structured, and how does it relate to answering correctly?
\end{list}

Each sub-question addresses a different part of the main question. RQ1 establishes whether the lecture context is sufficient to determine a poll's answer and to assign its instructional function, which provides the basis for the subsequent analysis. RQ2 uses this contextual reading to characterise how the instructors question the cohort. RQ3 turns to student responses, examining how engagement is distributed and whether it is associated with correctness. Throughout, engagement refers to behavioural engagement in the sense of Fredricks et al.~\cite{fredricks2004}, that is, whether a student responded; the design does not measure its emotional or cognitive components.

\subsection{Unit of Analysis and Coding}
\begin{figure}[htbp]\centering\includegraphics[width=0.8\textwidth,keepaspectratio]{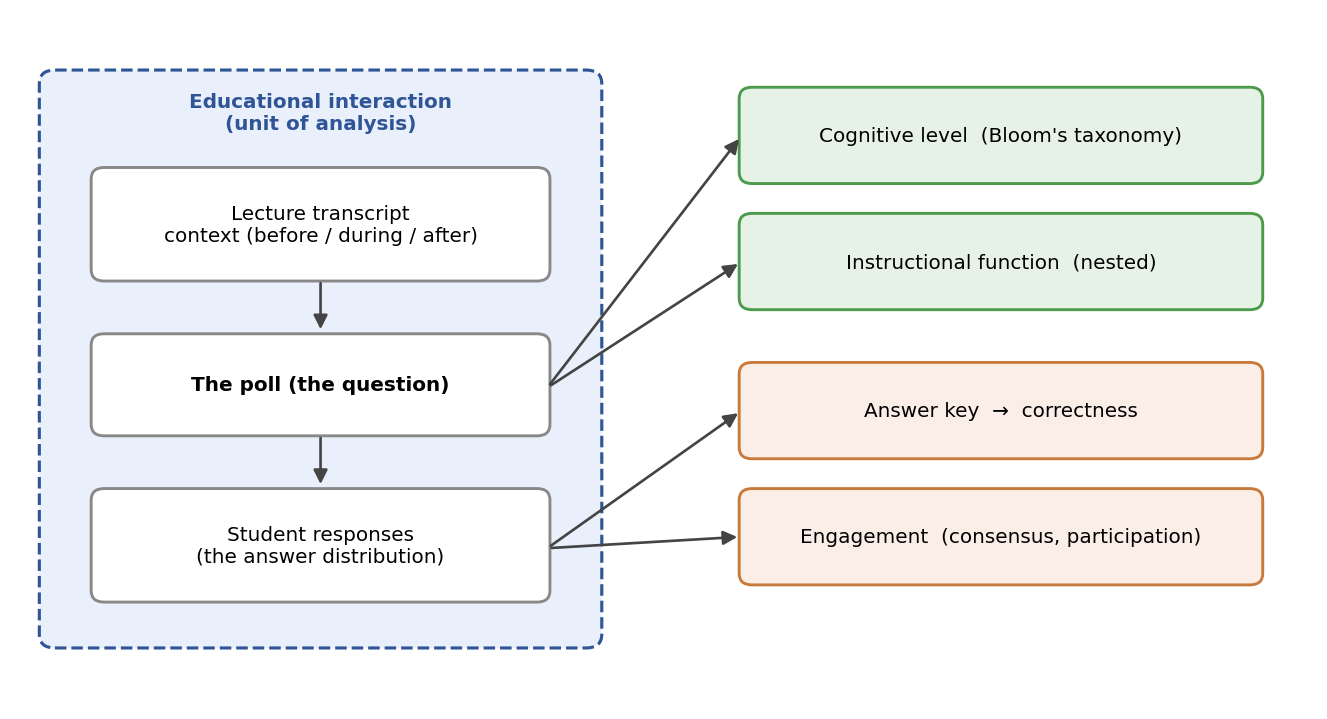}\caption{The poll as an educational interaction (the unit of analysis): its lecture context, the question, and the response distribution, from which we derive its cognitive level and instructional function, an answer key for correctness, and measures of engagement.}\label{fig:interaction}\end{figure}

We do not classify a poll only by its wording. Instead, we analyse the complete educational interaction surrounding the poll, including the lecture discourse before and after it, the poll question itself, and the distribution of student responses it draws (Fig.~\ref{fig:interaction}). Considering these elements together allows us to distinguish identical True/False items by the instructional function they serve.

For each interaction, we derive four types of information. First, we assign the poll a cognitive level based on the original Bloom's taxonomy. This classification reflects the cognitive process required of students in the instructional context, rather than the verbs used in the question or its format. Then in each cognitive level, we assign an instructional function that describes the purpose served by the poll. When the session establishes an objectively correct answer, we record it as the \emph{answer key} and use it to assess response accuracy. Polls addressing opinions, mood, or logistical matters are excluded from accuracy analysis because they do not have a correct answer. Finally, consensus, response spread, and participation rate are derived as three measures of engagement for students present when the poll was launched. Throughout the analysis, agreement and accuracy are treated as distinct measures.

\noindent\textbf{Coding procedure.} The coding was carried out with a large language model (a Claude Opus model, Anthropic) rather than by hand. For each poll, the model received the question together with the transcript from six minutes before the poll to three minutes after it and followed a written protocol: it first stated the instructional intent of the poll and the cognitive process expected of students, and only then assigned a Bloom level, recording a primary and an alternative level, a confidence rating, and a justification that cites the transcript. The intent recorded here is the intent the lecture context supports, not a claim about what the speaker privately intended. The model and the rule set had no access to the response distributions, so levels and functions were fixed independently of how students answered. Instructional functions were then assigned by a fixed rule set applied to the recorded intent and process; the seven functions are the categories this procedure produced, and we treat them as descriptive rather than as an established taxonomy. The coded output was reviewed by the authors, and the answer key was verified manually by an author for every poll on which the class majority disagreed with the derived answer, with 28 answers corrected. To assess reliability, an independent second automated coder relabelled a random sample of 40 cognitive polls: agreement with the rule-based functions was 65\% ($\kappa = 0.48$), with disagreements concentrated between adjacent functions such as Factual Recall and Attention/Distortion Check. Automated coding of educational text is an established practice, from classifiers that assign Bloom levels to questions \cite{mohammedomar2020} and automated measures of classroom discourse \cite{demszky2021} to language-model annotation whose agreement with human coders matches agreement among human coders in several tasks \cite{gilardi2023,xiao2023,tai2024}. The protocol, rule set, and reliability sample are given in the online appendix.

\subsection{Survey and Triangulation}
A short post-program student survey consisting of six five-point Likert-scale questions was conducted separately from the corpus analysis. Each question was designed to correspond to one measured finding, and the survey also included two open-ended questions about the perceived purpose of the polls and which polls students found difficult. We matched each respondent to their measured record by hashed identity and compared the two sources at the class-aggregate and per-student levels (Sect.~\ref{sec:triangulation}).

\section{Data Collection and Analysis}

\subsection{Study Context}
The corpus is drawn from 47 live sessions conducted across 39 calendar days as part of an online summer internship orientation program led by multiple speakers. The program focuses on motivation, professional culture, and exposure to concepts rather than graded mastery of skills. This \emph{orientation} genre is relevant to the interpretation of the results. It shapes the heavily lower-order questioning profile we report, so throughout we distinguish findings that are likely specific to this kind of program from those that should generalise to other settings.

\subsection{Data Streams}
Each session provides three forms of data that are time-aligned. The first is the live poll records, which contain the options presented for each poll, individual responses, and submission timestamps. The second is the lecture transcript, aligned with the session timeline, which records what was said at a particular time during the session. The third is attendance, recorded through per-attendee summaries and join and leave intervals for each learner, allowing us to determine who was present during each poll. The full corpus contains 604 poll questions, 340,668 responses, and 2,807 distinct learners. Most questions are True/False, and each session has a corresponding set of polls and transcript records.

\subsection{Alignment}
\begin{figure}[htbp]\centering\includegraphics[width=0.8\textwidth,keepaspectratio]{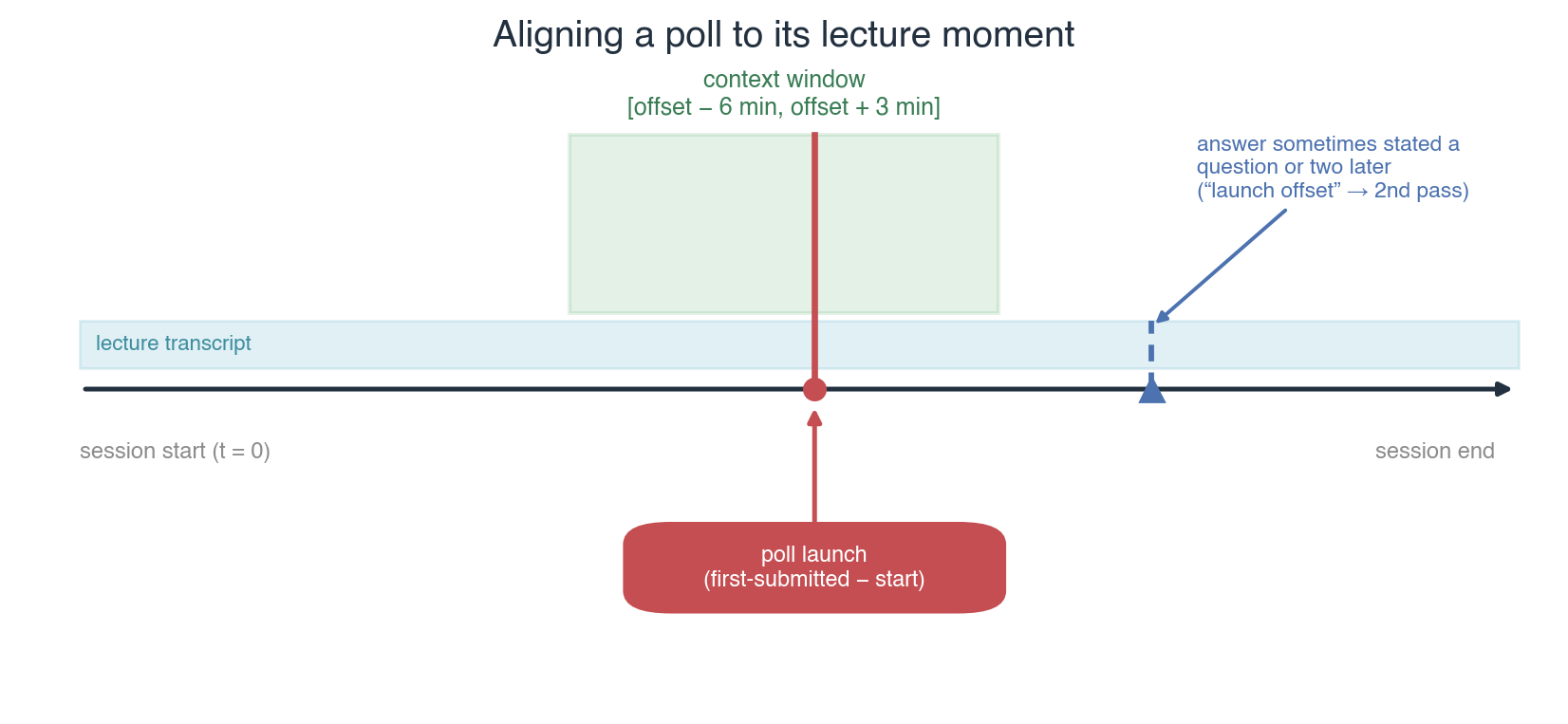}\caption{Reading a poll in its lecture moment: its launch time locates it on the transcript timeline, and the surrounding window supplies the talk that gives the question its meaning.}\label{fig:alignment}\end{figure}

To interpret the instructional function of each poll, we place the poll and transcript timestamps on a common timeline. For each poll, we calculate its session-relative time by subtracting the actual session start time from the timestamp of its first submitted response. We validated the resulting alignment against the timeline recorded in the transcript. We then examine the lecture content from roughly six minutes before to three minutes after the poll is launched. This window captures the content immediately before, during, and after the question and provides the context used to interpret its instructional function (Fig.~\ref{fig:alignment}).

\subsection{Data Handling and Ethics}
Prior to data collection, informed consent for the use of student data in the study was obtained from all participating students. The analysis uses only aggregate data or salted-hash pseudonyms, generated by applying a one-way hash to normalised email addresses. All cross-stream joins, including linking student poll responses with attendance records and later with survey responses, are performed using these pseudonyms. No individual student is identified in any table or figure. Continued participation in the program was partly contingent on answering polls. As a result, the recorded participation reflects both students' engagement and the requirement to answer the polls. We take this constraint into account when interpreting the results in Sect.~\ref{sec:results} (\emph{Results}).

\subsection{Supplementary Material}
\label{sec:supp}
The detailed summaries of the anonymised dataset, including corpus composition, the Bloom and instructional-function taxonomies, answer-key coverage and correctness, engagement measures, and a per-session table, are provided in an online appendix. It also includes the complete methods and results of the triangulation survey.\footnote{\url{https://osf.io/rxqhe/?view_only=48b1a715e1aa4cb598b341bd5939c7e2}}

\section{Results and Discussion}
\label{sec:results}

\subsection{Results}
\textbf{Reading a poll in its lecture context (RQ1).} The answer to 89\% of polls (539 of 604) could be objectively located in the surrounding lecture. The lecture context was therefore sufficient to settle what the correct response was for the large majority of polls. Reading the polls alongside their surrounding transcript also revealed a recurring type of attention-checking poll, the \emph{Attention/Distortion-Check}, in which a speaker restates a lecture point with a deliberate error, such as changing a number, inserting a negation, or attributing a claim incorrectly. In these cases, the statement may appear true when read on its own, while the surrounding lecture shows that it is false. The finding therefore shows that the context surrounding a poll can contain information about its instructional function that is not available from the question wording alone.

\textbf{Question types and instructional functions (RQ2).} We describe each cognitive poll in two ways: by its cognitive demand, using the six levels of the original Bloom's taxonomy, and by its instructional function, a finer category nested within each cognitive level. The two classifications capture different aspects of a poll. Cognitive demand describes the level of thinking a question requires, while instructional function describes the purpose the poll serves within the lecture. Together, they provide a fuller description of how questions are used across the cohort than either classification alone.

\begin{figure}[htbp]\centering\includegraphics[width=0.8\textwidth,keepaspectratio]{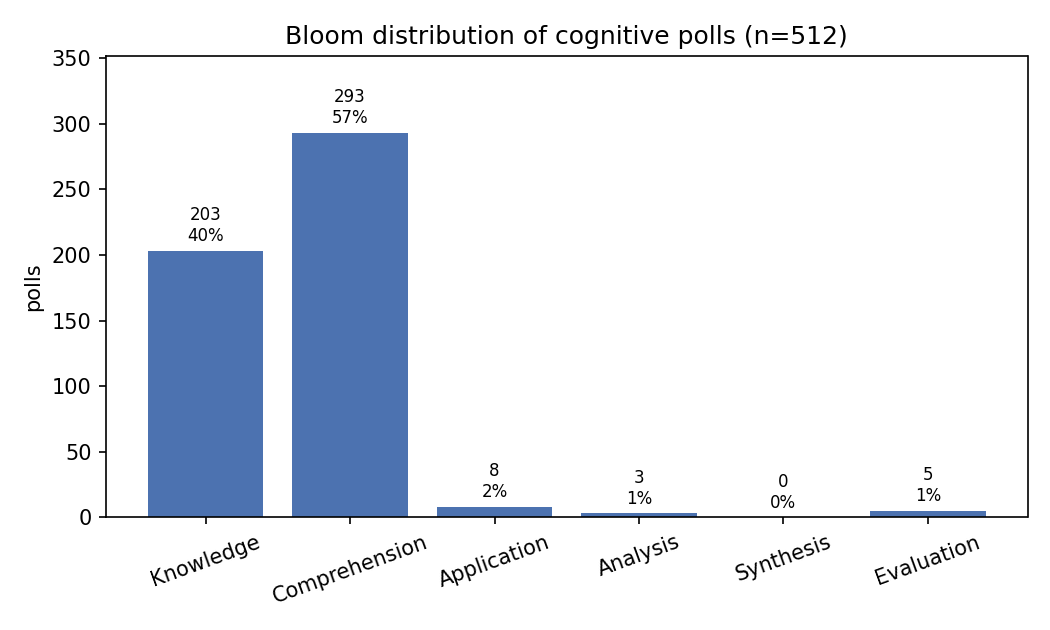}\caption{Cognitive demand of the 512 cognitive polls under Bloom's taxonomy: the questioning is concentrated at the Knowledge and Comprehension levels, with higher-order levels nearly absent.}\label{fig:bloom}\end{figure}

By cognitive demand, the polls are overwhelmingly lower-order. Of the 512 cognitive polls, 96.8\% fall at the Knowledge or Comprehension levels of Bloom's taxonomy (Fig.~\ref{fig:bloom}). Only sixteen polls are at higher-order levels, namely Application, Analysis, or Evaluation, while Synthesis is absent entirely. This distribution remains stable across the 39 days, with no clear shift toward more demanding questions as the program progresses. The pattern is consistent with the orientation format of the program, which emphasises exposure to ideas rather than graded mastery.

Nested within that lower-order band, the polls fall into seven recurring instructional functions (Table~\ref{tab:functions}). Four of these correspond to functions already described in the questioning and formative-assessment literature: checking recall \cite{roedigerkarpicke2006}, verifying comprehension of what was just taught \cite{chin2007,blackwiliam1998}, redirecting attention \cite{szpunar2013}, and applying a stated procedure \cite{redfieldrousseau1981}. The remaining three, reinforcing a stated principle, integrating ideas across the lecture, and a residual category, were needed to cover this corpus and are specific to it.

\begin{table}[htbp]
\centering
\caption{The seven instructional functions of the 512 cognitive polls, with their counts and share of cognitive polls.}
\label{tab:functions}
\small
\begin{tabular}{lrr}
\toprule
\textbf{Instructional function} & \textbf{$n$} & \textbf{\% of cognitive} \\
\midrule
Understanding Verification & 246 & 48.0 \\
Factual Recall & 129 & 25.2 \\
Attention / Distortion Check & 53 & 10.4 \\
Concept Reinforcement & 24 & 4.7 \\
Applied Execution & 22 & 4.3 \\
Concept Integration & 18 & 3.5 \\
Other Recall / Comprehension & 20 & 3.9 \\
\bottomrule
\end{tabular}
\end{table}

Two functions, verifying that a just-delivered idea has been understood and checking recall of a stated fact, together account for roughly three-quarters of the cognitive polls, while the attention-checking device introduced above is a small but distinctive category (10\%). Because each poll's function is assigned from the surrounding lecture context, we can ask whether function, rather than wording, is associated with how students respond. With the True/False format held constant, the spread of responses differs significantly across functions (Kruskal--Wallis $H = 17.8$, $p = 0.007$), whereas participation rate does not ($H = 4.7$, $p = 0.58$). This suggests that differences in how students respond are related to the instructional function of the poll and not to its surface format alone.

\begin{figure}[htbp]\centering\includegraphics[width=0.8\textwidth,keepaspectratio]{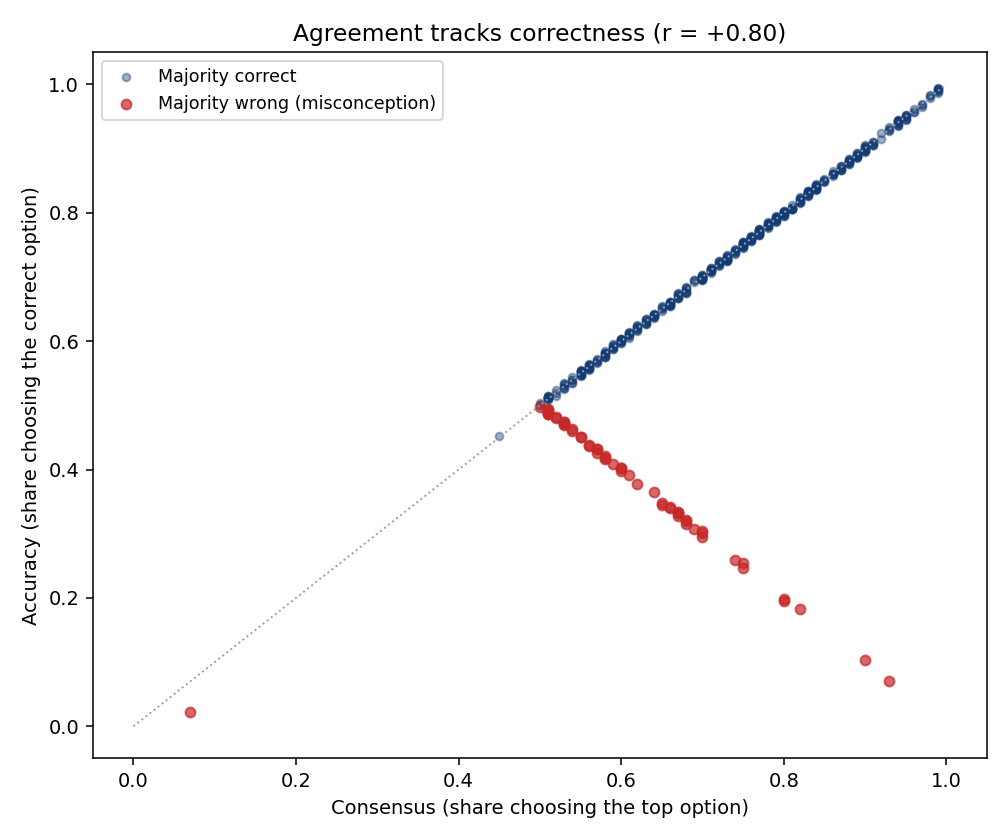}\caption{Consensus versus correctness across the decidable polls ($r \approx 0.80$); the confident misconceptions are the high-consensus yet incorrect exceptions.}\label{fig:consensus}\end{figure}

\textbf{Engagement structure and correctness (RQ3).} About 87\% of attendees answer at least one poll, yet participation is highly concentrated. The most active fifth of students account for 74\% of all responses, with a Gini coefficient of about 0.70. Against the answer key, the class majority answers correctly on 88.5\% of polls and incorrectly on 11.5\%. A small subset of these polls represent confident misconceptions, where a large majority selects the wrong answer. Such cases would be difficult to identify from agreement alone. Consensus and correctness are closely related ($r \approx 0.80$), as expected for binary items with a known answer, where a large majority is usually a correct one; the correlation is therefore largely a property of the format. The informative cases are the exceptions, the confident misconceptions that lie off this pattern (Fig.~\ref{fig:consensus}). The relationship between individual participation and answer accuracy is weak, indicating that participation alone provides limited information about understanding. Because continued participation in the program depended in part on answering polls, any apparent increase in engagement over time may also reflect a selection effect rather than genuine growth in engagement.

\begin{figure}[htbp]\centering\includegraphics[width=\textwidth,keepaspectratio]{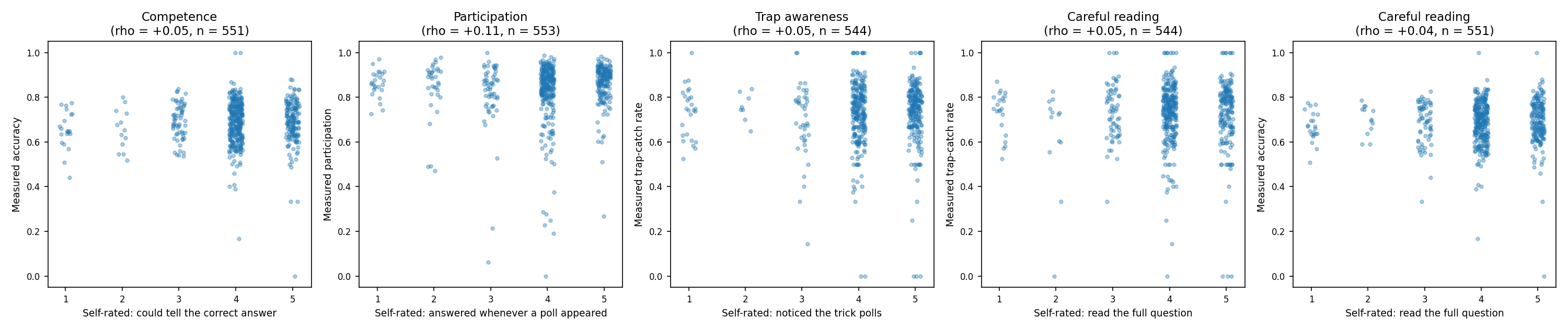}\caption{Self-report versus measured behaviour: only participation tracks its measure, while self-assessed competence and trap-catching do not.}\label{fig:triangulation}\end{figure}

\textbf{Triangulation against the student survey (RQ1--RQ3).}\label{sec:triangulation} An independent survey of 579 respondents, with 95\% matched to their measured records by hashed identity, provides a comparison with the corpus at two levels. At the class level, the two sources show similar patterns in how students perceive the polls. Students report that the polls are tied to the lecture (91\%), consistent with the 89\% of polls whose answers are locatable in the lecture context, and that the polls mainly test recall or basic understanding (84\%), consistent with the 96.8\% of cognitive polls classified at the Knowledge or Comprehension levels. At the individual level, self-reported participation tracks measured participation, reflecting a behaviour that students can directly observe. The two sources differ more clearly in measures of correctness. Students' self-assessed correctness does not track their measured accuracy, and among the 456 students who were confident that they could identify the answer, 49\% had accuracy below the median. Students also identify the attention-checking poll as both the most memorable and the most difficult, accounting for 46\% of difficulty responses, although these polls are not the most difficult to answer correctly (Fig.~\ref{fig:triangulation}). The survey therefore captures a difference between perceived difficulty and measured accuracy. Individual-level associations are further limited by a strong response ceiling and self-selection, since survey respondents are more accurate and more participatory than non-respondents. The results are interpreted as evidence of a gap between self-perception and measured performance rather than as evidence of no association. Full statistical details are provided in the online appendix (Sect.~\ref{sec:supp}).

\subsection{Discussion}
On their own, poll results primarily show correctness and participation. When these results are considered in the context of the lecture and compared with a second source, they provide a broader view of how a class is questioned and how students respond. Four observations support this interpretation.

\noindent\textbf{Context settles the answer, and it exposes the instructional function.} In 89\% of the polls, the answer could be objectively determined from the surrounding lecture context. Response distributions also varied according to the instructional function of the poll, while participation rates remained stable. Together, these findings indicate that the lecture context is sufficient both to settle a poll's answer and to assign its instructional function. They also suggest that differences in response patterns are associated with the instructional function of a poll rather than with its wording alone. The Attention/Distortion-Check illustrates what the wording alone would miss, as the statement appears true when read in isolation, but the surrounding lecture context shows that it is false.

\noindent\textbf{In this setting, polls function primarily as a participation activity rather than an assessment.} Cognitive demand is low and remains broadly stable, participation does not vary by instructional function, participation is broad, and individual participation is only weakly related to answer accuracy. Students also identify attention and engagement more often than assessment of understanding when describing the purpose of the polls. Taken together, these findings suggest that polling in this orientation program serves primarily to sustain participation rather than to assess mastery. As noted in Sect.~\ref{sec:results}, this is what the orientation format of the program would lead one to expect. Recovering correctness allows us to say so without dismissing the answers. The class answers correctly on roughly nine out of ten polls, so the response data carry a meaningful correctness signal, but that signal is a by-product of a participation activity rather than the outcome of a graded assessment.

\noindent\textbf{The highest-value signal is confident error.} Once correctness is recovered, a particularly informative pattern is agreement on an incorrect answer rather than disagreement. We refer to this as a confident misconception, where a large majority of students select the wrong answer. Such cases cannot be identified from agreement alone and become visible only when responses are compared with an answer key. This matters in practice because a live poll shows the instructor the distribution of responses, and a strong majority can easily be taken as a sign that the class has understood. Confident misconceptions are the cases in which that interpretation fails.

\noindent\textbf{Confidence is not competence.} The survey provides an individual-level view of this distinction. Students report the purpose of the polls and their own participation with reasonable consistency, but their confidence in whether they answered correctly does not track their measured accuracy. The polls that students report as most difficult are also not the polls on which they have the lowest accuracy. This gap between self-assessment and measured performance shows why objective, context-grounded measures are needed alongside student reports.

\subsection{Limitations}
There are several limitations to the claims made in this study. The corpus comes from a single orientation program, so the low and stable cognitive profile and the particular mix of instructional functions may reflect the nature of this program. The approach of reading polls in their lecture context, identifying their instructional function, and comparing the findings with survey responses can be applied in other settings, but the distributions observed here may differ. The survey is skewed toward agreement and represents a more engaged, self-selected group of students. This limits how we interpret the individual-level null results. We therefore treat these results as consistent with a gap between self-assessment and measured performance, rather than as evidence that no association exists. Answering polls was also partly required for continued participation in the program. The resulting participation measure reflects both student engagement and the requirement to answer polls. The participation requirement means that the participation data cannot show whether engagement increased or decreased because of the polls, nor can they establish an effect of polling on retention. The seven instructional functions identified in the corpus are specific to our analysis and should therefore be treated as an exploratory classification. The classifications are model-based with moderate second-coder agreement, so fine distinctions between adjacent functions should not be over-read. These points should be kept in mind when interpreting the findings, particularly when considering their relevance to other settings.

\section{Conclusion}

The classroom polls have the information about questioning and student engagement that is not captured by correctness scores or participation counts alone. Reading the polls in their lecture context allowed us to locate an objective answer for 89\% of them and to identify a recurring attention-checking function that could not be recognised from the poll wording alone. The response distributions varied while the participation rate remained stable for questions that were primarily lower-order and fell into seven instructional functions. Student engagement was broad but concentrated, and individual participation showed only a weak relationship with correctness. The answer key which was prepared from the lecture, showed that the majority of the class answered roughly nine out of ten polls correctly, while a small set of confident misconceptions emerged when agreement was considered alongside correctness. The survey findings were consistent with the corpus in their reports of poll purpose and participation, but students' self-assessed correctness did not fully correspond to their measured performance. Together, these findings show that the response record of a classroom poll contains information about the instructional event and student behaviour that is lost when polling is reduced to correctness and participation.

These findings gave insights for further studies. If we continue the study on the same cohort over time, it can help separate changes in participation from the effect of the program's participation rule. In the future, we can vary the function and timing of polls to examine how these factors relate to engagement and learning under controlled conditions. Applying the analysis across different instructors, subjects, and graded settings would show which parts of the observed questioning profile are specific to the orientation setting. The contextual approach could also be used during live sessions to examine how polls function and response patterns, including confident misconceptions, develop as the session progresses. This would help determine whether these measures can provide instructors with useful information during instruction.

\bibliographystyle{splncs04unsrt}

\bibliography{refs}

\end{document}